\documentclass[twocolumn]{aastex631}
\usepackage{amsmath}
\usepackage[T1]{fontenc}
\newcommand\bb[1]{\mbox{\boldmath{$#1$}}}
\usepackage{xcolor}
\usepackage{CJK}

\newcommand\revise[1]{#1}

\begin{document}

\title{Simple convective accretion flows (SCAFs):\\
Explaining the ${\approx-1}$ density scaling of hot accretion flows around compact accretors
}

\author[0000-0002-9408-2857]{Wenrui Xu \begin{CJK*}{UTF8}{gbsn}(许文睿)\end{CJK*}}
\affiliation{Center for Computational Astrophysics, Flatiron Institute, New York, USA}

\begin{abstract}
Recent simulations find that hot gas accretion onto compact accretors are often highly turbulent and diskless, and show power-law density profiles with slope $\alpha_\rho\approx-1$.
These results are consistent with observational constraints, but do not match existing self-similar solutions of radiatively inefficient accretion flows.
We develop a theory for this new class of accretion flows, which we dub simple convective accretion flows (SCAFs).
We use a set of hydrodynamic simulations to provide a minimalistic example of SCAFs, and develop an analytic theory to explain and predict key flow properties.
We demonstrate that the turbulence in the flow is driven locally by convection, and argue that radial momentum balance, together with an approximate up-down symmetry of convective turbulence, yields $\alpha_\rho=-1\pm~{\rm few}~0.1$. 
Empirically, for an adiabatic hydrodynamic flow with $\gamma\approx 5/3$, we get $\alpha_\rho\approx-0.8$; the resulting accretion rate (relative to the Bondi accretion rate), $\dot M/\dot M_{\rm B}\sim (r_{\rm acc}/r_{\rm B})^{0.7}$, agrees very well with the observed accretion rates in Sgr A*, M87*, and a number of wind-fed SgXBs.
We also argue that the properties of SCAFs are relatively insensitive to additional physical ingredients such as cooling and magnetic field; this explains its common appearance across simulations of different astrophysical systems.
\end{abstract}


\section{Introduction}\label{sec:intro}

From supermassive black holes (SMBHs) to compact objects around giant stars, many distinct astrophysical systems contain radiatively inefficient accretion flows onto small accretors whose size (or effective size) is many orders of magnitude below the characteristic outer scale of the flow (e.g., the Bondi radius). The large scale separation poses a challenge in simulating such systems, yet it provides opportunities for developing analytic, radially self-similar solutions that actual accretion flows may asymptote towards.
Beyond the simplest case of spherical Bondi accretion \citep{Bondi1952}, previous self-similar analytic theories for radiatively inefficient accretion flows usually consider a disk-like geometry (motivated by the large scale scale separation together with the assumption of angular momentum conservation, and by observational evidences of jets), with an overdense midplane and underdense poles; this contains a number of distinct solutions depending on whether the midplane contains an rotationally supported disk, whether the pole contains an outflow, and the level and origin of turbulence (see a review of some representative theories in \citealt{YuanNarayan2014}).

However, recent 3D simulations of radiatively inefficient accretion flows with large scale separations commonly demonstrate a class of flows whose properties are not well explained by existing theories.
Physically, this class of flows concern the accretion of hot (or radiatively inefficient; cf. \citealt{Shapiro1976,Ichimaru1977,Narayan1995}) gas at scales below the Bondi radius $r_{\rm B}$ (or a similar outer scale). They are characterized by two main features:
\begin{itemize}
    \item The flow is highly turbulent, and the mean flow properties are often \revise{closer} to spherical symmetry as opposed to disk-like;
    \item The radial mean density profile shows a power-law slope $\alpha_\rho\approx -1$. This leads to a power-law dependence of accretion rate on the accretor size $\dot M\propto r_{\rm acc}^{\alpha_{\dot M}}$ with slope $\alpha_{\dot M} \approx 0.5$.
\end{itemize}
Flows exhibiting these properties have been observed in a number of simulations that cover distinct physical systems such as accretion onto SMBHs like Sgr A* \citep{Pang2011,Ressler2018,Ressler2020a,Ressler2023} and M87* \citep{Guo2022}, wind accretion onto NSs in supergiant X-ray binaries (SgXBs; \citealt{XuStone2019}), and accretion during common envelope evolution \citep{MacLeod2015}.
Intriguingly, the $\rho$ and $\dot M$ scalings appear to be insensitive to the detailed physics of the flow, which varies significantly across some of these simulations.
They also seem to be broadly consistent with observational constraints on these systems, especially their low accretion rates relative to Bondi accretion. 
In the case of Sgr A* (cf. \citealt{Ressler2023} Fig. 3) and M87* \citep{Russell2015}, the radial density profiles can also be directly constrained from observations, and they show good agreement with $\alpha_\rho\approx -1$.
Furthermore, the $\alpha_\rho \approx -1$ scaling resembles that of a (not yet explained) empirical scaling seen in some simulations \citep[e.g.,][]{White2020, Ressler2020b, Begelman2022} of hot magnetically arrested disks (MADs), hinting that there might be some connection in between.

On the other hand, existing theories on radiatively inefficient accretion flows often fail to predict this $\alpha_\rho\approx-1$ scaling. When the scaling is set by constant mass flux, as in the case of Bondi accretion and advection-dominated accretion flows (ADAFs; \citealt{NarayanYi1994}), one gets $\alpha_\rho=-3/2$. When the scaling is set by constant energy flux, as in the case of convection-dominated accretion flows (CDAFs; \citealt{Narayan2000,QuataertGruzinov2000}), one gets $\alpha_\rho=-1/2$. \citet{Gruzinov2013} speculates that one could get $\alpha_\rho=-1$ if the flow consists of jets that conserve momentum, but whether such jets are produced in nature (or in simulations) remains an open question.

In order to provide a better theory for this new class of turbulent radiatively inefficient accretion flows, we introduce a new self-similar solution for radiatively inefficient flows whose properties are mainly set by strong convective turbulence.
We dub this new solution simple convective accretion flows (SCAFs).
The remainder of this paper aims to build a basic theoretical framework for SCAFs.
In Section \ref{sec:sim} we present numerical solutions of a toy problem that showcases the most basic form of SCAF in an adiabatic hydrodynamic flow.
In Section \ref{sec:theory} we develop a theory for SCAFs by demonstrating the convective nature of the turbulence and predicting analytically key flow properties, especially the density and accretion rate scalings.
We then discuss how and when these results can be generalized to different physical setups and compare SCAF against CDAF in Section \ref{sec:discussion}, show that SCAFs are consistent with a few observational constraints in Section \ref{sec:obs}, and summarize our findings in Section \ref{sec:summary}.

\section{A toy problem}\label{sec:sim}

As a basic example of SCAFs, we simulate an adiabatic hydrodynamic accretion flow with $\gamma=5/3$ onto a small accretor. We also perform simulations with a few other $\gamma$ values.
This toy model does not necessarily represent any realistic astrophysical accretion flow; we only use it to demonstrate the basic properties of SCAFs and show that these properties do not rely on additional physical ingredients or the injection of turbulence and/or rotation from larger scales.

\subsection{Simulation setup}
There is no dimensional free parameter in this problem, so we set $GM = r_0 = \rho_0 = 1$ where $M$ is the accretor mass, $r_0$ is the outer scale of the flow, and $\rho_0$ is the density at large radii. $r_0$ also sets a unit time of $t_0 = \sqrt{r_0^2/GM}$.

Our simulations are performed using the code \texttt{athena++} \citep{Stone2020} on a 3D Cartesian domain $x,y,z\in [-2r_0, 2r_0]$. 
The large-scale flow properties are set by a damping zone at $r>r_0$, where we damp density towards $\rho_0$, pressure towards $P_{\rm out}=\frac{\gamma-1}{\gamma}\frac{GM\rho_0}{r_0}$, and velocity towards zero, all with a rate of $t_0^{-1}$.
Here the choice of $P_{\rm out}$ corresponds to zero Bernoulli parameter (marginally bound); we tested that different choices of $P_{\rm out}$ would not affect flow properties at small scales.

We use a smooth sink region as the inner boundary condition. 
Within the sink (accretor) radius $r_{\rm acc}$, we remove mass (while conserving temperature and velocity) at a rate set by the dynamical timescale at $r_{\rm acc}$,
\begin{equation}
    \frac{{\rm d}\rho}{{\rm d}t} = -\sqrt{\frac{GM}{r_{\rm acc}^3}}\rho,
\end{equation}
and smooth the gravity to
\begin{equation}
    g_r(r) = \left\{\begin{array}{ll}
        -\frac{GM}{r^2} & (r>r_{\rm acc}) \\
        -\frac{GMr}{r_{\rm acc}^3} & (r\leq r_{\rm acc})
    \end{array}\right..
\end{equation}
Another common accretor setup in the literature is a vacuum sink, where density and pressure are set to small values (effectively zero) at every timestep.
We also perform a set of runs with a vacuum sink and compare the results in Section \ref{sec:toy:results}.

We use mesh refinement to increase resolution at small radii and maintain approximately constant angular resolution until the max level of refinement $l_{\rm max}$ is reached. The root grid resolution is $64^3$ cells for $x,y,z\in [-2r_0, 2r_0]$.
To examine how the scale separation $r_{\rm acc}/r_0$ affects the dynamics and accretion rate, we perform simulations at several different $r_{\rm acc}$ and $l_{\rm max}$.
We always choose $r_{\rm acc}$ and $l_{\rm max}$ so that $r_{\rm acc}$ corresponds to 4 cells at the highest level of refinement.

We start each set of simulations with uniform density $\rho_0$ and pressure $P_{\rm out}$ and a weak, turbulent velocity field. \revise{The turbulent velocity field contains a total kinetic energy of $0.1 GM\rho_0 r_0^2$, with a power spectrum $P_v \propto k^{-2}$ for wavenumbers (wavelengths per box size) between 0 and 16 in each direction. The exact choice here barely affects the result since this initial perturbation is only used to break the symmetry of the initial condition.} We gradually increase the resolution and decrease the accretor size across a set of runs, going from $l_{\rm max}=1$ to 5, 8, 11, and 14. The highest-resolution ($l_{\rm max}=14$) run corresponds to $r_{\rm acc}/r_0=3.1\times 10^{-5}$.
Each run uses the end state of the previous run as the initial condition.
For runs with lower resolution ($l_{\rm max}=1,5,8$), we run for $100~t_0$.
For runs with higher resolution ($l_{\rm max}=11, 14$), we run for $\sim 10^6~t_{\rm acc}$ where $t_{\rm acc}\equiv\sqrt{r_{\rm acc}^3/GM}$ is the dynamical timescale at $r_{\rm acc}$.

\subsection{Simulation results}\label{sec:toy:results}

\begin{figure*}
    \centering
    \includegraphics[scale=0.6]{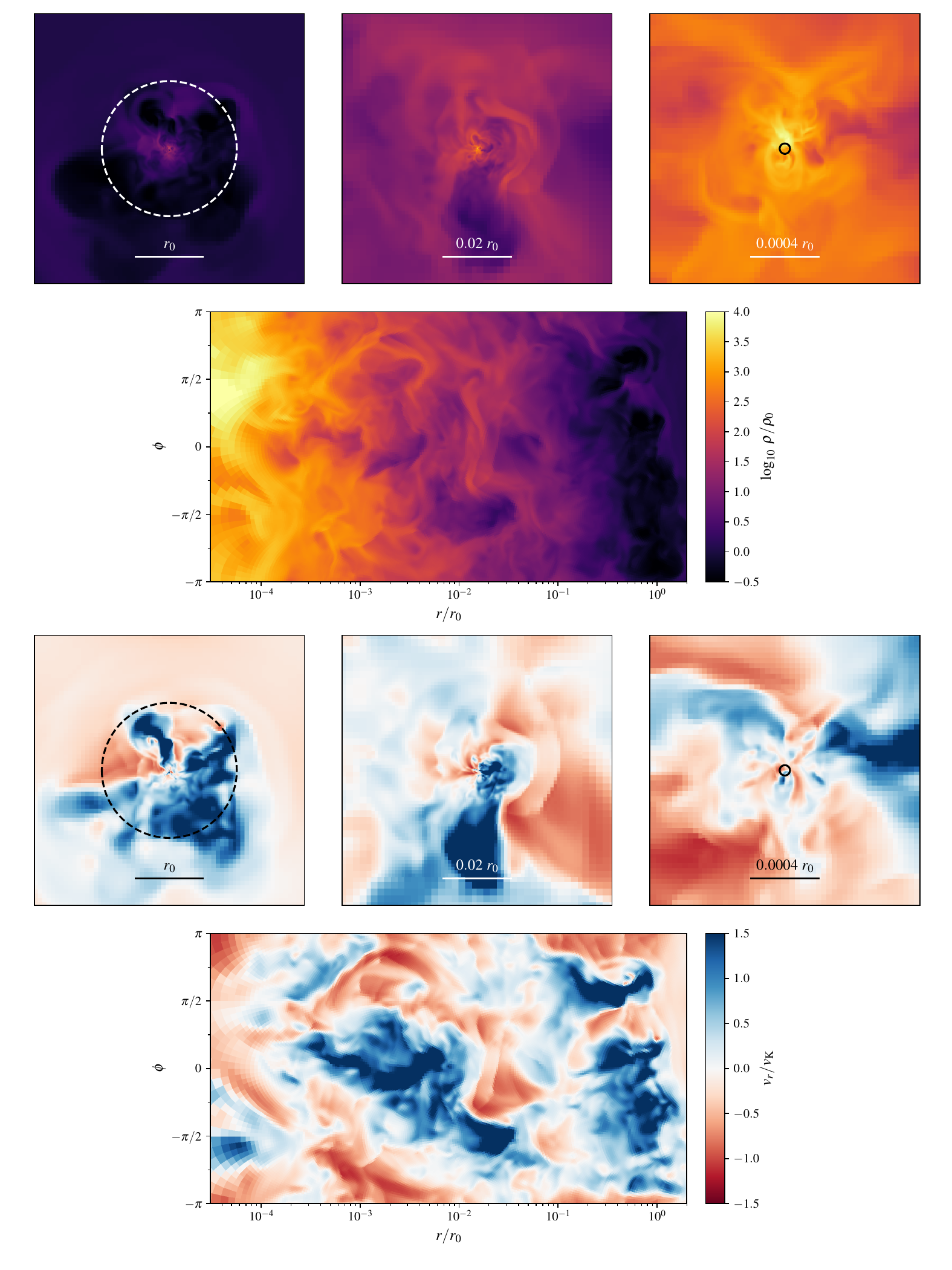}
    \caption{A snapshot taken from our fiducial simulation ($\gamma=5/3$, smooth sink) with $r_{\rm acc}=3.1\times 10^{-5}r_0$ ($l_{\rm max}=14$). First row: A slice at $z=0$ showing density profile at different scales; the circles mark $r=r_0$ and $r=r_{\rm acc}$, respectively. Each panel is a $50\times$ zoom-in of the previous one. Second row: The same slice mapped into polar coordinates. Last two rows: Same as the first two rows, but showing the radial velocity normalized by $v_{\rm K}\equiv \sqrt{GM/r}$. The flow shows strong, large-scale turbulence.}
    \label{fig:snapshot}
\end{figure*}

\begin{figure*}
    \centering
    \includegraphics[scale=0.66]{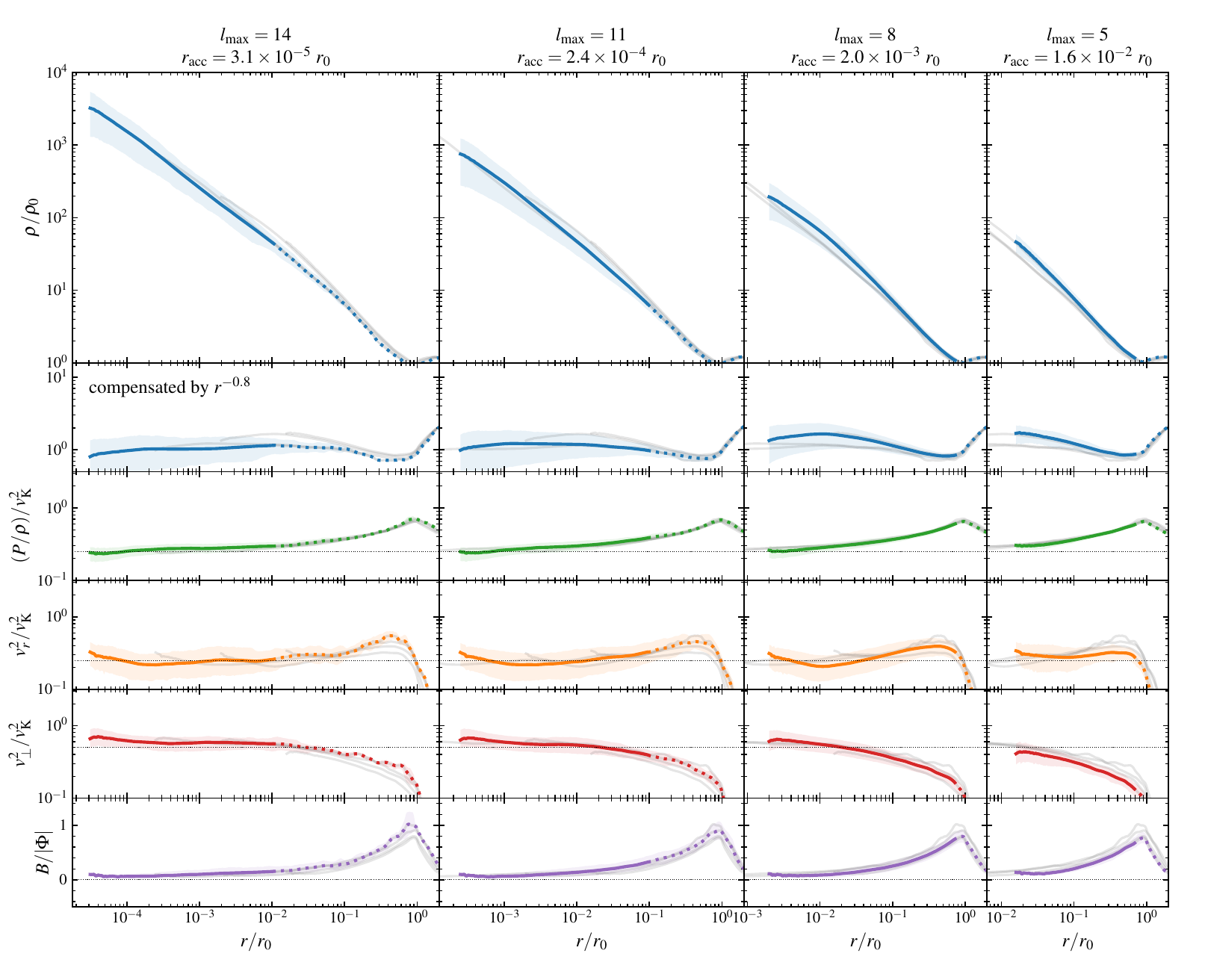}
    \caption{Radial profile of our fiducial simulations ($\gamma=5/3$, smooth sink). Each column shows results from a different $r_{\rm acc}$, and results from other $r_{\rm acc}$ are plotted in light grey lines for comparison. The rows show density $\rho$, density compensated by $r^{-0.8}$, temperature $P/\rho$, radial velocity $v_r^2$, tangential velocity $v_\perp^2$, and the normalized Bernoulli parameter $B/|\Phi|$. Here the Bernoulli parameter $B$ is defined as $B=E_{\rm tot}+P+\Phi$, with $E_{\rm tot}$ being the total (kinetic + thermal) energy and $\Phi=-GM\rho/r$ being the gravitational energy. Temperature and velocities are normalized by $v_{\rm K}^2$. Shaded area shows the variability of the flow, with the shade spanning between 16th and 84th percentiles.
    Solid lines mark regions in steady state, which we define as where the integration time is longer than $100\times$ the local dynamical timescale $r/v_{r,\rm rms}$. The profiles are relatively insensitive to $r_{\rm acc}$ and becomes approximately self-similar at small radii.}
    \label{fig:radial_profile}
\end{figure*}

\begin{figure}
    \centering
    \includegraphics[scale=0.66]{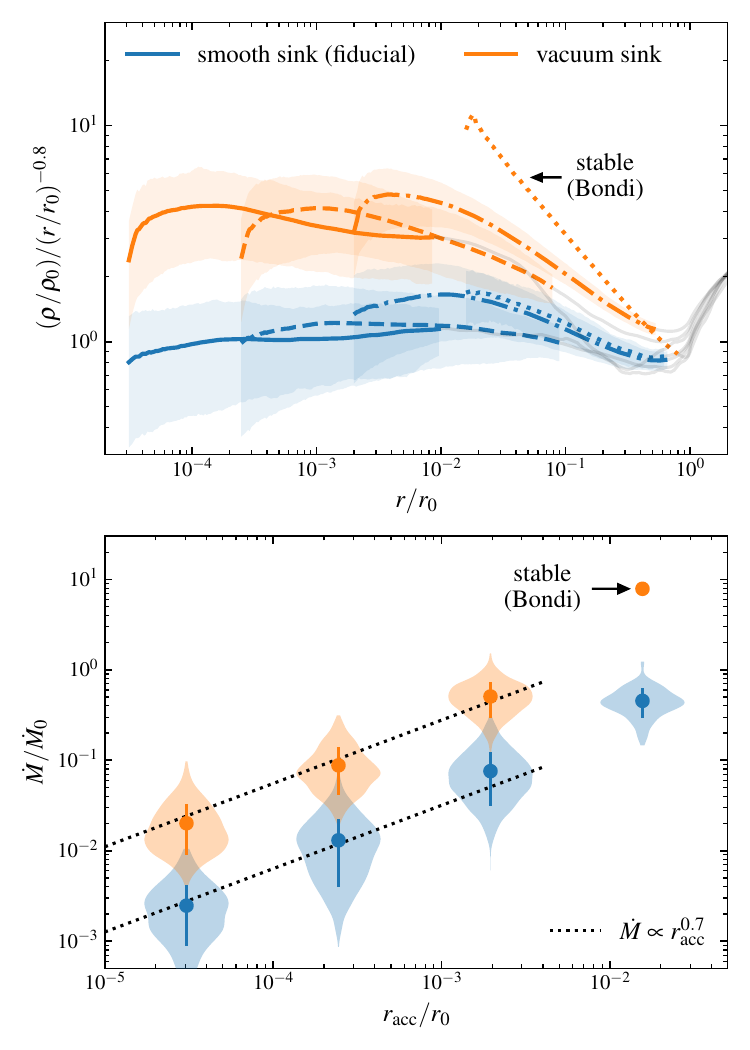}
    \caption{Comparison between runs with different inner boundary conditions (marked by colors) and different $r_{\rm acc}$.
    Top panel: compensated radial density profiles similar to those in Fig. \ref{fig:radial_profile}.
    \revise{Colored and grey lines show the part of the flow that has and has not reached steady state, respectively;} runs with different $r_{\rm acc}$ are marked by different line styles.
    For small $r_{\rm acc}$, we see a tendency of convergence towards $\alpha_\rho\approx -0.8$.
    Bottom panel: accretion rates. Dots show the mean accretion rate, error bars the 16th and 84th percentiles, and shaded areas the distribution. For small $r_{\rm acc}$, the accretion rate is consistent with a power-law scaling of $\alpha_{\dot M}\approx 0.7$.
    }
    \label{fig:Mdot_and_rho_scaling}
\end{figure}

\begin{figure}
    \centering
    \includegraphics[scale=0.66]{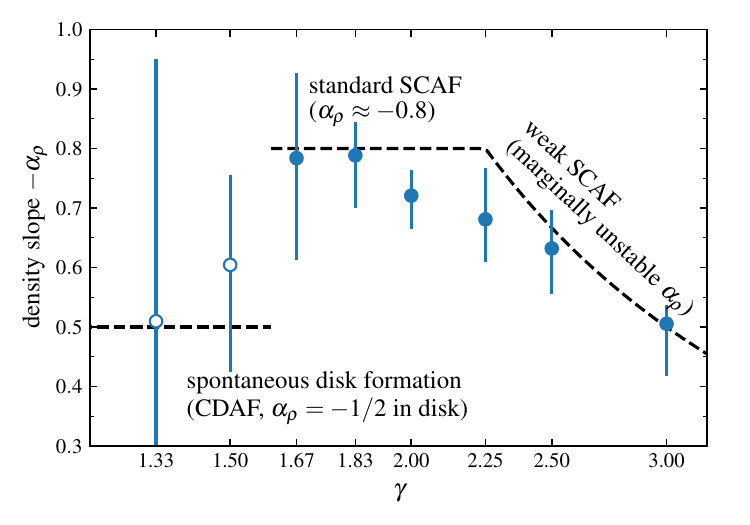}
    \caption{Slope of the self-similar density profile $-\alpha_\rho$ for different choices of the adiabatic index $\gamma$. All runs in this figure have $r_{\rm acc}= 2\times 10^{-3}r_0$ ($l_{\rm max}=8$). There are three main regimes. At intermediate values of $\gamma$, we see behavior similar to our fiducial $\gamma=5/3$ runs, with approximately constant $\alpha_\rho$.
    At large $\gamma$, turbulence is weaker and $\alpha_\rho$ is set by marginal convective instability, $\alpha_\rho = -1/(\gamma-1)$ (Section \ref{sec:dis:highgamma}). At small $\gamma$ (empty markers), a small disk resembling a CDAF forms spontaneously around the accretor (Section \ref{sec:dis:disk}).
    The slopes are estimated using the mean density profile for $r/r_{\rm acc}\in [1,20]$ for runs exhibiting SCAFs (solid markers) and $r/r_{\rm acc}\in [1,5]$ for runs exhibiting spontaneous disk formation (empty markers; this range corresponds to the self-similar disk portion of the flow). Markers and error bars show the median and 16th/84th percentiles of local density slope ${\rm d}\log\bar\rho/{\rm d}\log r$, respectively.
    }
    \label{fig:gamma_dependence}
\end{figure}

An overview of our fiducial simulations with $\gamma=5/3$ and a smooth sink is shown in Figs. \ref{fig:snapshot} and \ref{fig:radial_profile}.
The flow exhibits strong, large-scale turbulence, with turbulent velocity at the order of the Keplerian velocity $v_{\rm K}\equiv \sqrt{GM/r}$ and correlation length comparable to $r$ (Fig. \ref{fig:snapshot}). There is no organized rotation or outflow.

The flow is approximately self-similar, with kinetic, thermal, and gravitational energy comparable to one another and a power-law density profile (Fig. \ref{fig:radial_profile} and Fig. \ref{fig:Mdot_and_rho_scaling} top panel)
\begin{equation}
    \rho \propto r^{\alpha_\rho},~{\rm with}~\alpha_\rho \approx -0.8.
\end{equation}
In Section \ref{sec:scaling} we explain the origin of this scaling.
This power-law density profile leads to a power-law dependence of the accretion rate on the accretor size (Fig. \ref{fig:Mdot_and_rho_scaling}, bottom panel),
\begin{equation}
    \dot M \propto r_{\rm acc}^{\alpha_{\dot M}},~{\rm with}~\alpha_{\dot M} = \frac 32 + \alpha_\rho \approx 0.7.
\end{equation}
Here we have assumed $v_r\propto v_{\rm K}\propto r^{-1/2}$ at $r\sim r_{\rm acc}$.

Fig. \ref{fig:Mdot_and_rho_scaling} also shows a comparison between the smooth sink runs and the vacuum sink runs.
At $r_{\rm acc}\gtrsim 10^{-2}~r_0$, the accretion flow with a vacuum sink is stable, probably because a large vacuum sink removes turbulent energy more efficiently.
For smaller $r_{\rm acc}$, vacuum sink runs are qualitatively similar to smooth sink runs, although they show slightly steeper density profiles and higher density and accretion rates.\footnote{The slightly steeper $\alpha_\rho$ for vacuum sinks at finite $r_{\rm acc}/r_0$ is probably one reason why previous studies (e.g., \citealt{Ressler2018,XuStone2019,Guo2022}; all of which adopt vacuum sinks) tend to report $\alpha_\rho = -1$ and $\alpha_{\dot M} = -0.5$.}
The difference in the density slope $\alpha_\rho$ decreases towards smaller $r$, suggesting that runs with different boundary conditions asymptote towards the same self-similar solution. \revise{Unlike the slope, the intercept of the density profile does not seem to converge towards the same value. The exact reason for this dependence of density intercept on the sink properties remains unclear, but it probably hints a feedback mechanism that allows the dynamics near the sink to affect dynamics at the outer scales (cf. Section \ref{sec:bernoulli}).}

We also perform a set of runs with different values of $\gamma$ ranging from $4/3$ to 3. Although many $\gamma$ values here may not directly correspond to any astrophysical system, this serves as a simple example of how varying the underlying physics of the flow impact flow properties.
The resulting $\alpha_\rho$ is summarized in Fig. \ref{fig:gamma_dependence}.
For $1.5\lesssim \gamma \lesssim 2$, the flow properties and $\alpha_\rho$ are largely similar to $\gamma=5/3$.
For $\gamma\gtrsim 2$, the flow properties are still qualitatively similar, except turbulence tends to be weaker and $\alpha_\rho$ shallower; this regime will be discussed in Section \ref{sec:dis:highgamma}.
For $\gamma\lesssim 1.5$, a disk with $\alpha_\rho\approx -1/2$ (resembling a CDAF) spontaneously forms around the accretor; this regime will be discussed in Section \ref{sec:dis:disk}.

\section{A theory for SCAFs}\label{sec:theory}

\subsection{What drives the turbulence?}\label{sec:turbulence}
\begin{figure}
    \centering
    \includegraphics[scale=0.66]{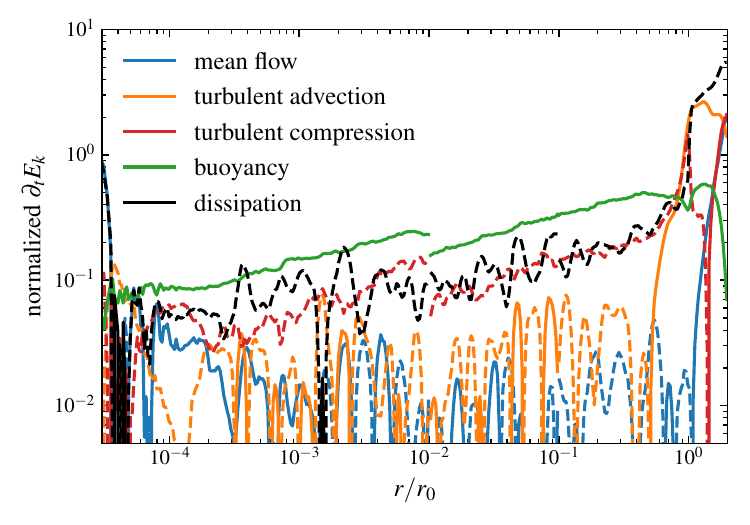}
    \caption{Contribution of different sources to the kinetic energy evolution $\partial_t E_k$ (with $E_k\equiv\frac 12\overline{\rho v^2}$), normalized by $E_k \Omega_{\rm K}$. Each line corresponds to a term in Eq. \eqref{eq:Ek}, and are marked with different colors. Solid/dashed lines represent positive/negative contributions. Buoyancy (convection) is the main source of kinetic energy. This figure combines data from three simulations: $r_{\rm acc}\approx 3\times 10^{-5} r_0$ ($l_{\rm max}=14$) for $r/r_0 < 10^{-2}$, $r_{\rm acc}\approx 2\times 10^{-4} r_0$ ($l_{\rm max}=11$) for $10^{-2}<r/r_0 < 10^{-1}$, and $r_{\rm acc}\approx 2\times 10^{-3} r_0$ ($l_{\rm max}=8$) for $r/r_0 > 10^{-1}$. \revise{This produces minor discontinuities at the intersections.}}
    \label{fig:convection}
\end{figure}

Explaining the properties of SCAFs first requires identifying the origin of the turbulent motion. Below we demonstrate that the turbulence in our simulations is excited locally by convection, contradicting previous conjectures that the turbulence is driven by outflows originating around the accretor \citep{Gruzinov2013,XuStone2019}. \revise{We comment that this subsection only discusses how the turbulence is sustained against dissipation in steady state. Meanwhile, a different mechanism may be accountable for triggering the turbulence (i.e., destabilizing the laminar Bondi flow); we discuss this later in Section \ref{sec:dis:highgamma}.}

A simple estimate of convective stability is given by the Brunt-V\"ais\"al\"a frequency,
\begin{equation}
    N^2 = -\frac{1}{\rho}\frac{{\rm d}P}{{\rm d}r}\left(\frac{1}{\gamma}\frac{{\rm d}\ln P}{{\rm d} r} - \frac{{\rm d}\ln \rho}{{\rm d} r}\right).\label{eq:brunt}
\end{equation}
For a small adiabatic radial displacement, $-N^2$ gives the ratio between buoyant acceleration and displacement.
For a self-similar flow with $\gamma=5/3$ and $\alpha_\rho\approx -1$,
\begin{equation}
    N^2 = \frac{(1-\alpha_\rho)P}{\rho r^2}\frac{(1-\gamma)\alpha_\rho - 1}{\gamma} < 0,
    \label{eq:Nsq_ss}
\end{equation}
and we expect the flow to be unstable.

It is worth noting that in a strong turbulence, the buoyant acceleration may not be the only important source of radial acceleration; in particular, when angular momentum is conserved, the centrifugal acceleration acts to stabilize radial motion into an epicyclic oscillation.
However, there are a couple of reasons why this epicyclic effect would not suppress convection altogether.
First, turbulent tangential motion may break angular momentum conservation quickly via scattering between parcels, causing the $v_\perp$ of a displaced parcel of fluid to be determined mainly by the typical $v_\perp$ at its current location instead of its original angular momentum.
Second, even if the radial motion is linearly stable, the finite level of mixing (in entropy and/or momentum) still causes a correlation between displacement and radial velocity, which then allows the buoyant acceleration to inject kinetic energy. (Meanwhile, the centrifugal force does not remove kinetic energy; it only coverts radial and tangential kinetic energy to each other.)

In addition to the qualitative arguments above, we test quantitatively whether convection is the main source of kinetic energy in our simulations.
Using the Euler density and momentum equations, we get the equation for kinetic energy evolution
\begin{equation}
    \partial_t\left(\frac 12 \rho v^2\right) = -\nabla\cdot\left(\frac 12 \rho v^2 \bb{v}\right) - \bb{v}\cdot \nabla P + \rho \bb{v}\cdot\bb{g} - \epsilon.
\end{equation}
\revise{Here $\epsilon$ captures the dissipation by irreversible processes (i.e., shocks and turbulent cascades); $\epsilon$ is formally zero in smooth, inviscid hydrodynamics}.
Averaging in $\theta,\phi$ and time at a given radius and separating the mean ($\bar\rho$ etc.) and turbulent ($\rho'=\rho-\bar\rho$ etc.) components, we get
\begin{equation}
\begin{split}
    \partial_t\left(\frac 12 \overline{\rho v^2}\right) =& \underbrace{-\nabla\cdot\left(\frac 12 \overline{\rho v^2} \bb{v}_0\right) - \bb{v}_0\cdot \nabla \bar{P} + \bar{\rho} \bb{v}_0\cdot\bb{g}}_{\text{mean flow}}\\
    &\underbrace{-\nabla\cdot\left(\frac 12 \overline{(\rho v^2) \bb{v}_1}\right)}_{\text{turbulent advection}}
    \underbrace{- \overline{\bb{v}_1\cdot \nabla P'}}_{\substack{{\rm turbulent}\\{\rm compression}}}
    \underbrace{- \overline{\bb{v}_1}\cdot \nabla \bar{P}}_{\text{buoyancy}}\\
    &\underbrace{- \epsilon}_{\text{dissipation}}.
    \label{eq:Ek}
\end{split}
\end{equation}
Here we have defined mean and turbulent velocity $\bb{v}_0, \bb{v}_1$ as
\begin{equation}
    \bb v_0 \equiv {\overline{\rho\bb{v}}}/{\bar \rho},~~~\bb v_1 \equiv \bb v - \bb v_0.
\end{equation}
Note that in general $\overline{\bb v_1}\neq 0$. Additionally, in the limit of low accretion rate, $\bb v_0\to 0$ and $\bb v_1\approx \bb v$.

In Eq. \eqref{eq:Ek} we label the physical significance of each term.
If dynamics near the accretor affects larger scales by launching outflows and sound waves, these effects will be captured in the turbulent advection term and the turbulent compression term.
The term $-\overline{\bb{v}_1}\cdot \nabla \bar{P}$ represents work done by buoyancy because one can rewrite it as
\begin{equation}
    - \overline{\bb v_1} \cdot \nabla \bar{P} = \overline{\rho \bb v \cdot\underbrace{\left[\left(\frac{1}{\rho}-\frac{1}{\bar\rho}\right)(-\nabla \bar{P})\right]}_{\text{buoyant acceleration}}}.
\end{equation}
In simulations, the dissipation term $\epsilon$ corresponds to the dissipation at grid scale, and we obtain $\epsilon$ by directly evaluating all other terms in Eq. \eqref{eq:Ek}.

In Fig. \ref{fig:convection} we compare the terms in Eq. \eqref{eq:Ek} and find that the buoyancy term is indeed the main source of kinetic energy, confirming that turbulence is driven mainly by convection.

\subsection{Explaining the density and accretion rate scaling}\label{sec:scaling}

As demonstrated in our toy problem and many previous simulations, SCAFs show the following characteristic radial scalings of density and accretion rate,
\begin{align}
    &\alpha_\rho \equiv \frac{{\rm d}\log\rho}{{\rm d}\log r} \approx -1,\\
    &\alpha_{\dot M} \equiv \frac{{\rm d}\log\dot M}{{\rm d}\log r_{\rm acc}} \approx 0.5.
\end{align}
In this subsection we discuss how the radial momentum balance of the flow sets these scalings.

To summarize, our argument is that in a flow with strong turbulence, $\alpha_\rho+1$ is comparable to the ratio between mean and typical acceleration in the flow; this ratio is then approximately zero because the turbulence, with zero net mass flux, exhibits approximate up-down symmetry.\footnote{
\revise{Here ``up'' and ``down'' are defined with respect to the direction of gravity (up/down refers to radially outward/inward). This notation follows the literature of convective flows in stars, which bares some resemblance to the turbulent spherical flow in our problem.}
}
Below we present this argument in detail.

Consider the radial component of the momentum equation,
\begin{equation}
    -r^{-2}\partial_r(r^2 \overline{\rho v_r^2}) - \partial_r \bar{P} + r^{-1}\overline{\rho v_\perp^2} + \bar\rho g_r = 0.
    \label{eq:momentum_balance}
\end{equation}
Here $\bb v_\perp$ is the tangential ($\theta,\phi$) component of velocity.
The latter terms in Eq. \eqref{eq:momentum_balance} correspond to the average radial acceleration (including centrifugal acceleration),
\begin{equation}
    \bar\rho a_{r,0} = - \partial_r \bar{P} + r^{-1}\overline{\rho v_\perp^2} + \bar\rho g_r.
    \label{eq:a_r}
\end{equation}
Combining Eqs. \eqref{eq:momentum_balance} and \eqref{eq:a_r}, we get
\begin{equation}
    \bar\rho a_{r,0} = r^{-2}\partial_r(r^2 \overline{\rho v_r^2}) = (1+\alpha_\rho) \frac{\overline{\rho v_r^2}}{r},
    \label{eq:alpha_in_a_r_1}
\end{equation}
\begin{equation}
    \alpha_\rho = - 1 + \frac{\bar\rho a_{r,0}}{\overline{\rho v_r^2}/r}.
    \label{eq:alpha_in_a_r}
\end{equation}
Eq. \eqref{eq:alpha_in_a_r} relates $\alpha_\rho$ to the mean acceleration $a_{r,0}$. We comment that this result assumes strong turbulence (\revise{${v_r^2}/v_{\rm K}^2$ is a finite constant to lowest order}); otherwise Eq. \eqref{eq:alpha_in_a_r_1} is trivially satisfied (for this regime, see Section \ref{sec:dis:highgamma} and \ref{sec:dis:disk}).

Now consider the second term in Eq. \eqref{eq:alpha_in_a_r}. For a strong convective turbulence, the mixing length $\xi$ is comparable to $r$ (cf. \citealt{Arnett2018}) and ${v_r^2}/r$ represents the typical acceleration of the flow,
\begin{equation}
a_{\rm typical} \sim \frac{{v_r^2}}{\xi} \sim \frac{{v_r^2}}{r}.
\end{equation}
Meanwhile, for a self-similar flow with spherically symmetric mean flow properties and $\alpha_\rho\neq -3/2$ ($\alpha_\rho=-3/2$ would instead correspond to Bondi accretion or ADAF), the mean mass flux needs to be zero in the limit of $r_{\rm acc}\ll r\ll r_0$. The turbulent motion thus has an approximate up-down symmetry: there are equal amount of inward and outward turbulent motion, and that requires approximately equal inward and outward turbulent acceleration. Cancellation between inward and outward acceleration leaves the mean acceleration much smaller than the typical acceleration,
\begin{equation}
|a_{r,0}|\ll a_{\rm typical} \sim \frac{{v_r^2}}{r}.
\end{equation}
This implies
\begin{equation}
    |a_{r,0}| \ll \frac{{v_r^2}}{r},~~~|\alpha_\rho + 1|\ll 1.
\end{equation}
We comment that $|a_{r,0}|$ is generally not exactly zero and $\alpha_\rho$ not exactly $-1$ because the up-down symmetry of the turbulence is only approximate when $\xi/r$ is finite.

In Appendix \ref{a:finite_xi} we estimate the amplitude of $|a_{r,0}|$ from a different (and more quantitative) perspective to show that for $\xi\sim r$ the amplitude of $|\alpha_\rho+1|$ is at most few 0.1; this is also consistent with our empirical findings in simulations (Fig. \ref{fig:gamma_dependence}).
However, it is difficult to give a more accurate analytic estimate of $\alpha_\rho$. Modeling convection from first principle is known to be a challenging problem, and the large-scale ($\xi\sim r$) convective turbulence in SCAFs only increases this difficulty.

The radial slope of $\rho$ can then be easily translated to the $r_{\rm acc}$ dependence of $\dot M$.
Since the flow transitions into an infall with $v_r\sim -v_K$ near the accretor, the accretion rate is simply
\begin{equation}
    \dot M \sim r_{\rm acc}^2 \rho(r_{\rm acc})v_{\rm K}(r_{\rm acc}) \propto r_{\rm acc}^{\frac 32+\alpha_\rho},
\end{equation}
\begin{equation}
    \alpha_{\dot M} = \frac 32 + \alpha_\rho \approx 0.5.
\end{equation}

\subsection{Energy flux and Bernoulli parameter}\label{sec:bernoulli}

The radial scaling of $\alpha_\rho\approx -1$ also constrains the energy flux of the system.
In a steady-state radiatively inefficient flow, the energy flux averaged in time and on spherical shells $F(r)$ is constant in radius. (Here we only consider the non-radiative energy flux, since radiation is decoupled from the flow; the radiative energy flux will be discussed later in this subsection.)
Meanwhile, self similarity gives a characteristic dynamical energy flux of
\begin{equation}
F_{\rm dyn}(r) \equiv r^2 \bar\rho v_{\rm K}^3 \propto r^{\alpha_\rho+1/2},
\end{equation}
which increases towards smaller radii.
This together requires that $F(r)/F_{\rm dyn}(r)=0$ to zeroth order of $r/r_0$. Note that $F(r)$ itself is constant and generally finite, \revise{and can be large enough to affect the dynamics near the outer scale. The finite and constant $F(r)$ thus allows some coupling between the innermost and outermost scales of the system, which might explain why in Fig. \ref{fig:Mdot_and_rho_scaling} the intercept of the density profile (set by the non self-similar region at $r\sim r_0$) seem to depend on the sink properties even for $r_{\rm acc}\ll r_0$.}

In the special case of a hydrodynamic SCAF, the requirement of zero energy flux (compared to $F_{\rm dyn}$) implies zero Bernoulli parameter (compare to the characteristic dynamical energy density $|\Phi|$).
The energy equation can be written as
\begin{equation}
    -r^{-2}\partial_r(r^2\overline{\rho b v_r}) = 0,
\end{equation}
where the specific Bernoulli parameter $b$ is defined by
\begin{equation}
    \rho b \equiv E_{\rm tot} + P + \Phi = \frac 12 \rho v^2 + \frac{\gamma}{\gamma-1}P - \frac{GM}{r}\rho.
\end{equation}
Because there is no net mass flux, the energy flux $\overline{\rho b v_r}$ only comes from the difference between $b$ of outward and inward parcels ($b_{\rm out},b_{\rm in}$). Since adiabatic displacement conserves $b$, this difference mainly comes from the difference in mean specific Bernoulli parameter ($b_0$) at the original location of these parcels,\footnote{In general, $b$ is also affected by buoyant acceleration and mixing with surrounding material. Here we do not consider these mechanisms because their effects are the same (to lowest order) on $b_{\rm out}$ and $b_{\rm in}$ and cancel out in the difference $b_{\rm out}-b_{\rm in}$.} and to lowest order in the mixing length $\xi$ we have
\begin{equation}
    b_{\rm out}-b_{\rm in} \sim -\frac{{\rm d} b_0}{{\rm d} r}\xi.
\end{equation}
Therefore the energy flux is
\begin{equation}
    \overline{\rho b v_r} \sim -\frac 12 \overline{\rho |v_r|} \frac{{\rm d} b_0}{{\rm d} r}\xi.
\end{equation}
In self similarity, $b_0\propto |\phi| \propto r^{-1}$, as the gravitational potential $\phi\equiv-GM/r$ gives the characteristic scale of specific energy. Therefore, $F/F_{\rm dyn}=0$ requires that
\begin{equation}
    b_0/|\phi| = 0.
\end{equation}
i.e., the normalized Bernoulli parameter $B/|\Phi|$ in a hydrodynamic SCAF is approximately zero. This is also demonstrated in our simulations (Fig. \ref{fig:radial_profile}) and in \citet[][Fig. 8]{Guo2022}.

While the non-radiative energy flux in a SCAF is small (compared to $F_{\rm dyn}$), the radiative energy flux could be much larger because the accretion luminosity is
\begin{equation}
    F_{\rm rad} \sim \frac{GM\dot M}{r_{\rm acc}} \sim \left(\frac{r_{\rm acc}}{r_0}\right)^{\alpha_{\dot M} - 1}F_{\rm dyn}(r_0) \gg F_{\rm dyn}(r_0).
\end{equation}
This large radiative energy flux, powered by accretion, would not interfere with the gas dynamics when the flow is radiatively inefficient with negligible opacity, but could affect the dynamics at larger scales where radiative heating is no longer negligible.

\section{Generalization and discussion}\label{sec:discussion}
\subsection{Robusity of SCAFs} \label{sec:discussion:robusity}
In reality, accretion flows around compact objects are often not as simple as the $\gamma=5/3$ adiabatic hydrodynamic flow in our toy problem.
It is therefore important to understand whether our results remain applicable when we modify the physics of the flow.

Fundamentally, the scalings of $\rho$ and $\dot M$ discussed in Section \ref{sec:scaling} only rely on the following assumptions:
\begin{itemize}
    \item The flow is self-similar and in steady state, with strong turbulence ($v_r^2\sim v_{\rm K}^2$, $\xi\sim r$);
    \item The turbulence is driven locally by convection or a similar instability;
    \item Accretion is inefficient, with $v_{r, 0}/v_{\rm K} \to 0$ (zero mean mass flux) in all directions.
\end{itemize}
These assumptions are fairly generic, thus we expect the scalings to be robust against certain modifications to the physics of the flow, such as
\begin{itemize}
    \item Deviation from exact adiabaticity or $\gamma=5/3$, as long as it does not fully suppress convection at $\alpha_\rho\approx -1$ or cause disk formation (see later discussion in Section \ref{sec:dis:highgamma} and \ref{sec:dis:disk});
    \item Deviation from spherical symmetry (e.g., having a denser midplane and/or coherent rotation);
    \item Having a region that does not satisfy some of the assumptions, provided that the angular size of this region is small (or finite but constant) and the mass and radial momentum flux through the boundary of this region are negligible;
    \item Inclusion of a magnetic field (with or without mean field and/or net flux) that is also self-similar and satisfies the other assumptions (in particular, the magnetic field should not fully suppress instability or launch a high-$\dot M$ outflow).
\end{itemize}

The last several points deserve some more detailed explanation. When we relax the assumption that time-averaged flow properties are spherically symmetric and include possible contribution from magnetic field, integrating the radial momentum equation over all directions (solid angle $\Omega$) still yields
\begin{equation}
    -r^2\partial_r\left(\int r^2\overline{\rho v_r^2}{\rm d}\Omega\right) + \int \bar\rho a_{r,0} {\rm d}\Omega = 0.
    \label{eq:force_general}
\end{equation}
Here the averaging ($\bar\rho$, $a_{r,0}$, etc.) is done only in time but not in $\Omega$. The acceleration $a_{r,0}$ includes contribution from pressure gradient, centrifugal acceleration, gravity, and magnetic pressure and tension. If accretion is inefficient (zero mean mass flux) in all directions, we can still argue that $a_{r,0}\approx 0$ in all directions (with the argument in Section \ref{sec:scaling} or Appendix \ref{a:finite_xi}), and Eq. \eqref{eq:force_general} yields $\alpha_\rho\approx -1$. Therefore, deviation from spherical symmetry or inclusion of magnetic field does not directly affect the radial scalings of SCAFs. 
One important caveat is that it is no longer obvious whether the mean mass flux needs to be zero in all directions; it is reasonable to have self-similar (but not spherically symmetric) flows with finite positive $v_{r,0}$ in some directions and finite negative $v_{r,0}$ in others.

Meanwhile, if the flow contains a small region where our assumptions for SCAFs do not hold, we can exclude this region from the integration in Eq. \eqref{eq:force_general} and add a term for the radial momentum flux across the boundary of this region. As long as this term is small and the excavated region has negligible or constant angular size, we still get $\alpha_\rho\approx -1$. One example of this scenario would be a collimated, high-velocity jet that barely have time to interact with the rest of the flow before propagating further out.
This makes it possible for SCAFs to be applicable to realistic SMBHs, which may launch powerful jets that can dominate the energy flux at $r\gg r_{\rm acc}$ but are collimated and take relatively little volume \citep[e.g.,][]{Lalakos2022}.

The robusity of the radial scalings of SCAFs have been demonstrated in several simulations. In our toy model, small variation of $\gamma$ leaves $\alpha_\rho$ approximately constant (Fig. \ref{fig:gamma_dependence}), suggesting that the flow is not too sensitive to the exact level of convective instability.

\citet{Ressler2018,Ressler2020a} perform hydrodynamic and magnetohydrodynamic simulations of wind accretion onto Sgr A* and find that accretion flows with and without magnetic field show very similar scalings even though the magnetic field is dynamically important in the magnetized simulation. This is consistent with our argument that a self-similar magnetic field (in this case produced mainly by turbulent dynamo) would not affect the scalings.

\citet{Guo2022} also find $\alpha_{\dot M}$ of a multiphase accretion flow to be insensitive to whether cooling is included. In their simulations, accretion at small scale ($\ll$ Bondi radius) is dominated by the approximately adiabatic hot gas.
Meanwhile, although cooling produces cold gas that dominates mass and angular momentum at small radii, for most of the time the cold gas forms a rotationally supported disk which accretes very inefficiently.
Their results might be interpreted as an approximately non-accreting cold disk placed inside a hot SCAF; it is therefore not too surprising that whether this cold disk exists barely affects the accretion rate scaling.
One caveat though is that the hot gas does seem less turbulent when the cold disk is present, calling into question whether it is turbulent enough to be considered as a SCAF.

\subsection{Potential connections with hot MADs}\label{sec:MAD}

Following the arguments above, we further conjecture that hot MADs might be interpreted as SCAFs fueled by a slightly different instability. \citet{Begelman2022} recently argue that the strong turbulence in hot MADs is in fact driven by a convection-like instability. This makes it possible to apply the argument of SCAF scalings to the disk portion of hot MADs, provided that the outflow does not interact with the disk too much. Additionally, if the poloidal magnetic flux is transported by the turbulence passively as conjectured in \citet{Begelman2022}, one might argue that the up-down symmetry (or, more precisely, in-out symmetry) of magnetic flux transport also requires $a_{r,0}\approx 0$ and $\alpha_\rho\approx -1$.

Indeed, several MAD simulations with relatively thick (hot) disks and/or relatively weak outflows (Simulation C in \citealt{White2020}; \citealt{Ressler2020b,Ressler2021}; \citealt{Begelman2022}) show radial density profiles with $\alpha_\rho\approx -1$, whereas simulations showing thinner disks and stronger outflows have different radial scalings (Simulation A and B in \citealt{White2020}).
Furthermore, the simulation in \citet{Ressler2020b} shows the transition between a SCAF at larger scales and a MAD at smaller scales, and the spherically averaged density profile shows a clean, uninterrupted $\alpha_\rho\approx -1$ power law across the two regimes (see their Fig. B1), hinting that this profile may be set by fundamentally the same physics in both regimes.

\subsection{The effect of suppressing convection}\label{sec:dis:highgamma}

Since our theory predicts that a SCAF should have $\alpha_\rho\approx -1$, a natural question to ask is what happens when convection is suppressed so that a $\alpha_\rho\approx -1$ flow is still convectively stable. We explore this regime using simulations with large $\gamma$ (Fig. \ref{fig:gamma_dependence}).
Here the large values of $\gamma$ do not necessary resemble any real flow; we only use it to increase the stability of the flow, which in reality could be due to mechanisms such as cooling and magnetic field.

Suppressing convection does cause the flow to deviate from the $\alpha_\rho\approx -1$ scaling, yet the flow remains turbulent. We find $\alpha_\rho$ to be broadly consistent with marginal convective instability, which requires (cf. last term in Eq. \ref{eq:Nsq_ss})
\begin{equation}
    \alpha_\rho = -\frac{1}{\gamma-1}.
\end{equation}

It seems somewhat puzzling why turbulence persists even when we suppress convection. One possibility is that the turbulence is triggered fundamentally by a global instability: The small size of the accretor, together with (approximate) angular momentum conservation, causes any small asymmetry at large scale to suppress accretion and trigger turbulence at small scale \revise{-- a scenario theorized long ago by \citet{Shvartsman1971} and more recently demonstrated in simulations \citep[e.g.,][]{XuStone2019}.}
Then, under the requirement that the flow has to be turbulent, the only allowed self-similar state is $\alpha_\rho\approx -1$ for strong convection and $\alpha_\rho\approx -1/(\gamma-1)$ for weak convection.
Following this argument, a non-turbulent flow is achievable only when some other mechanism (e.g., magnetic braking) removes angular momentum as material moves inward.

\subsection{Comparison with CDAF and spontaneous disk formation}\label{sec:dis:disk}

\begin{figure*}
    \includegraphics[scale=.66]{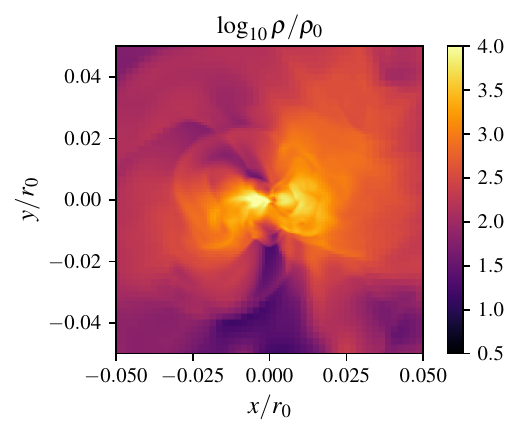}
    \includegraphics[scale=.66]{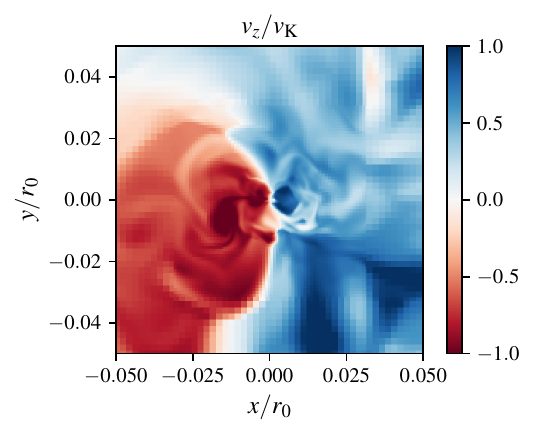}
    \includegraphics[scale=.66]{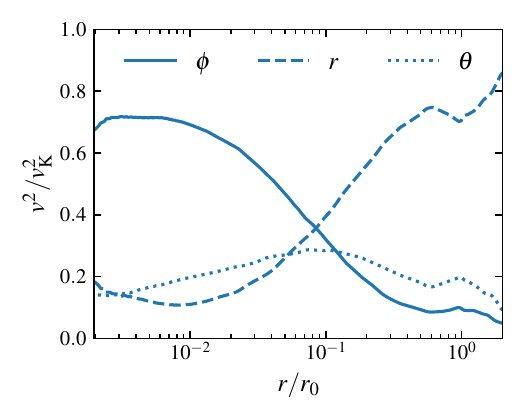}
    \caption{Left and center: density and azimuthal velocity of a snapshot taken from our $\gamma=4/3$ run with $r_{\rm acc} = 2\times 10^{-3}~r_0$ ($l_{\rm max}=8$), showing a disk/torus resembling a CDAF at the center. We pick an epoch when the disk angular momentum is approximately in $y$ direction. Right: flow properties in disk coordinate, averaged on spherical surfaces and in time. Here the $r,\theta,\phi$ directions are defined separately for each snapshot and each radius, with the pole aligned with the total angular momentum within this radius. At $r\lesssim$ few $10^{-2}~r_0$, the flow is supported by rotation (together with pressure) and the velocity is mainly in the azimuthal ($\phi$) direction. The amplitude of the azimuthal velocity agrees well with the CDAF theory, which predicts $v_\phi^2/v_{\rm K}^2 = 0.77$ for $\gamma=4/3$ \citep[][Eq. A15]{QuataertGruzinov2000}}
    \label{fig:g43}
\end{figure*}

SCAF is not the only solution for convective self-similar radiatively inefficient accretion flows. Another such solution is convective dominated accretion flow (CDAF; \citealt{Narayan2000,QuataertGruzinov2000}), which shows good agreement with 2D simulations and yield different scalings. In particular, CDAF has $\alpha_\rho = -1/2$ while SCAF has $\alpha_\rho\approx -1$.

The different properties of CDAF and SCAF are due to different underlying physical assumptions. The fundamental assumption of CDAF is that convection produces inward angular-momentum transport. The flow is thus disk-like (with $\bar v_\phi\propto v_{\rm K}$) and only marginally convective when other sources of outward angular-momentum transport (e.g., viscous transport) are weak or absent. The requirement of constant energy flux sets $\alpha_\rho$, and marginal convection sets other flow properties (cf. \citealt{QuataertGruzinov2000}). SCAF, on the other hand, takes the opposite assumption of high amplitude convection (turbulent $v_r\sim v_{\rm K}$). This assumption implicitly requires that convection either transports angular momentum outward (which suppresses disk formation as in our simulations) or is not the only source of angular-momentum transport (which might correspond to MADs). This different assumption affects what sets the flow properties. The radial momentum balance now sets $\alpha_\rho$ from Eq. \eqref{eq:alpha_in_a_r_1} (Section \ref{sec:scaling}). [Eq. \eqref{eq:alpha_in_a_r_1} is trivially satisfied in CDAFs where $v_r^2\approx 0$.] Meanwhile, since $\alpha_\rho\neq -1/2$, the energy flux has to be zero to lowest order, and this places a constraint on other flow properties (Section \ref{sec:bernoulli}).

The comparison above suggests that whether a given flow corresponds to CDAF or SCAF depends on the details of angular-momentum transport: when convection produces more inward/outward angular momentum transport, the flow tends to become a CDAF/SCAF.

Both CDAFs and SCAFs appear in our simulations. Runs with larger $\gamma$ produce SCAFs, while runs with $\gamma\leq 1.5$ spontaneously form disks that resemble CDAFs (Figs. \ref{fig:gamma_dependence} and \ref{fig:g43}). This trend agrees with previous simulations showing that increased compressibility (due to small $\gamma$ or cooling) promotes disk formation \citep{MacLeod2017,ElMellah2019,XuStone2019}. We also comment that our results and these previous studies together suggest that the physics of the flow, such as compressibility, is the main factor determining whether a disk can be formed and whether the flow becomes CDAF or SCAF. The existence of large-scale organized rotation and angular-momentum injection, on the other hand, appears to be less important. For example, in \citet{XuStone2019}, $\gamma=5/3$ flows cannot form disks even when there is angular-momentum injection from large scale. Meanwhile, in our simulations, $\gamma\leq 1.5$ flows form disks spontaneously without any large-scale angular-momentum injection. 

Another factor that might be important is the Rayleigh number, which characterizes the ratio between convection and dissipation. It is defined as ${\rm Ra} = -N^2 L^4 / \chi\nu$ where $N^2$ is the Brunt-V\"ais\"al\"a frequency, $L$ the outer scale of the turbulence, $\chi$ the thermal diffusivity and $\nu$ the kinematic viscosity. \citet{LesurOgilvie2010} find that as the Rayleigh number increases (stronger convection and/or weaker dissipation), the direction of angular-momentum transport changes from inward to outward. One interesting implication of this result is that sometimes the same flow (same $L,\chi,\nu$) might allow both CDAF (small $-N^2$, inward angular-momentum transport) and SCAF (large $-N^2$, outward angular-momentum transport).

Finally, whether the flow prefers CDAF or SCAF could also be affected by numerical factors, in particular resolution (as the level of numerical dissipation affects the effective Rayleigh number) and the dimensionality of the simulation. For $\gamma=5/3$, axisymmetric 2D simulations (e.g., \citealt{Stone1999}, run K) produce CDAFs while 3D simulations (cf. Section \ref{sec:intro}) generally produce SCAFs. This is probably because the assumption of axisymmetry in 2D simulations eliminates the component of the Reynolds stress produced by azimuthal variation of $v_r$ and $v_\phi$, thereby reducing the outward angular-momentum transport by turbulence.

In summary, CDAF and SCAF correspond to two distinct regimes of convective accretion flows. Empirically, 3D simulations often find SCAFs, but it is difficult to predict analytically whether a given flow corresponds to CDAF or SCAF because the details of angular-momentum transport depends on a number of not well-understood physical and numerical factors. We also comment that since angular-momentum transport (either inward or outward) by hydrodynamic convective turbulence is generally weak \citep[e.g.,][]{StoneBalbus1996}, the distinction between CDAF and SCAF in real astrophysical systems is probably set mainly by other physics such as the magnetic field.

\section{Consistency with observation}\label{sec:obs}

\begin{figure}
    \centering
    \includegraphics[scale=0.66]{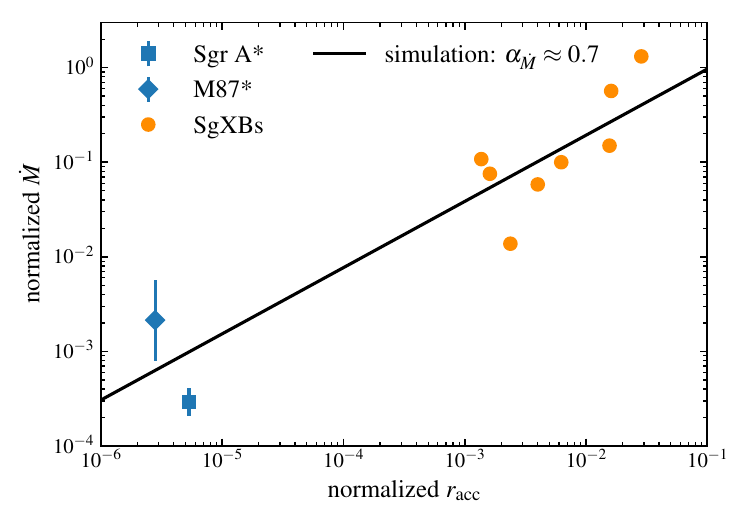}
    \caption{Observational estimates of normalized accretor size and accretion rate in two SMBHs and a number of wind-fed SgXBs, compared against our empirical accretion rate scaling $\alpha_{\dot M}\approx 0.7$ (based on $\gamma=5/3$ simulations; black line). Our simulation result shows good agreement with observation.
    Here, observational estimates of $r_{\rm acc}$ and $\dot M$ are normalized by the characteristic radius and accretion rate of Bondi or Bondi-Hoyle-Lyttleton accretion.
    Details of these estimates are presented in Appendix \ref{a:obs}.
    For Sgr A* and M87*, the error bars show uncertainty in the accretion rate, which is the main source of uncertainty. For SgXBs, we do not show error bars because the uncertainties are difficult to quantify.
    }
    \label{fig:obs_Mdot}
\end{figure}

\begin{figure}
    \centering
    \includegraphics[scale=0.66]{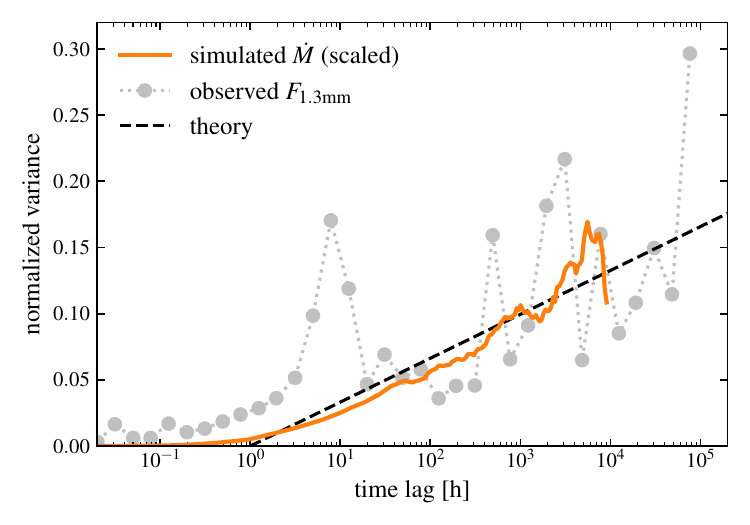}
    \caption{
    Structure function of the 1.3 mm luminosity of Sgr A* (grey, taken from \citealt{Dexter2014}), in comparison with the theoretical expectation for SCAFs, which is a log-linear relation between time lag and variance (black dashed line) at larger $\Delta t$. We also show the structure function of $\dot M$ from our $l_{\rm max}=14$ vacuum sink simulation (orange) for comparison; we scale it by a factor of $0.18$ to fit observation. Physically, allowing this extra factor is reasonable because the slope of the structure function could depend on the detailed physics of the flow (e.g., magnetic field, cooling) and, more importantly, the 1.3 mm luminosity is generally not a linear function of $\dot M$ as the spectrum of emission can shift with $\dot M$ \citep[cf.][]{Witzel2018}.
    }
    \label{fig:obs_SgrA}
\end{figure}

Several pieces of direct and indirect observational evidence suggest that the density, accretion rate, and variability scaling of hot accretion flows in nature are consistent with the scalings of SCAFs.

X-ray observation of M87* near its Bondi radius with resolution comparable to $R_{\rm B}$ \citep{Russell2015} suggests a $\alpha_\rho\sim -1$ density scaling for $\sim 1-$few $R_{\rm B}$. They also give a density estimate of $n_e\sim 0.3~{\rm cm}^{-3}$ at $\sim 0.1~{\rm kpc}$. More recently, \citet{EHT2021} give order-magnitude constraints on the plasma properties at $\sim 5 R_{\rm g}$ scale, obtaining $n_e\sim 10^{4-7}~{\rm cm}^{-3}$ from a one-zone model and $n_e\sim 10^{4-5}~{\rm cm}^{-3}$ from simulations. The density estimates at $\sim R_{\rm B}$ and $\sim R_{\rm g}$ together suggest a slope of $\alpha_\rho \sim 0.94-1.15$, broadly consistent with SCAFs. (Note that variability of the flow can cause the slope at any given epoch to show finite deviation from the mean slope.)
Similar results are found for Sgr A* from a set of independent measurements; see a summary in \citet[][Fig. 3]{Ressler2023}.

Previous studies also find that $\alpha_{\dot M}\approx 0.5$ is broadly consistent with observed accretion rates of Sgr A* \citep{Ressler2018}, M87* \citep{Guo2022}, and a number of SgXB systems \citep{XuStone2019}.
To make a more direct comparison across different systems, we summarize the accretor size $r_{\rm acc}$ and accretion rate $\dot M$ of these systems normalized by their characteristic (Bondi or Bondi-Hoyle-Lyttleton) accretion radius and accretion rate in Fig. \ref{fig:obs_Mdot}.
The observations, across $\sim 4$ orders of magnitude in $r_{\rm acc}$, show remarkable agreement with our $\gamma=5/3$ simulation result of $\alpha_{\dot M}\approx 0.7$.

The accretor's variability can also be used as a sanity check for SCAF. For emission coming from $\sim r_{\rm acc}$, there are two main sources of variability: a short-term variability due to the dynamics in and near the emission region, and a long-term variability due to the dynamics far from the accretor, which causes fluctuations of the mean density near the accretor. If the accretor is fed by a SCAF, the latter source of variability should be a pink noise. In terms of the structure function ${\rm SF}^2$, which is the variance of flux density difference as given time lag, this gives (as a general feature of self-similar local variability; cf. \citealt{Lyubarskii1997})
\begin{equation}
    \frac{{\rm d} {\rm SF}^2(\Delta t)}{{\rm d}\log \Delta t} \approx {\rm constant}.
\end{equation}
Indeed, such structure function shows up in both our toy model and in observation of Sgr A* (Fig. \ref{fig:obs_SgrA}). One caveat is that the observed structure function at large $\Delta t$ is very noisy and may contain artifacts due to irregular sampling. Therefore, determining whether the scaling is log-linear (pink noise) or flat (white noise) may require more detailed full light curve modeling in the future.

It is also worth noting that since SCAFs and (hot) MADs have very similar scalings (cf. Section \ref{sec:MAD}), they are both consistent with the observations discussed above. For SMBH accretion, a SCAF+MAD scenario where the flow transitions from an approximately spherical SCAF at larger radii to a MAD at smaller radii might be a more realistic description, as shown by simulations \citep[e.g.,][]{Ressler2020b,Lalakos2022}.

\section{Summary}\label{sec:summary}

We discuss a new class of self-similar solutions describing the turbulent accretion of hot gas around compact objects, simple convective accretion flows (SCAFs).
SCAFs exhibit strong, large-scale turbulence driven locally by convection (or a similar instability), and show power-law radial density profiles that result in low, accretor-size-dependent accretion rates.
SCAFs have been observed in numerical simulations for sometimes very different physical setups (cf. Section \ref{sec:intro}), yet until now there lacks a clear explanation for what drives the turbulence, what sets the radial scalings, and why similar scalings persists when additional physical ingredients (e.g., cooling and magnetic field) are included.
In order to answer these questions, we carry out numerical experiments for the simplest version of SCAFs (Section \ref{sec:sim}) and develop a series of analytic argument to explain and predict the flow properties (Sections \ref{sec:theory} and \ref{sec:discussion}). We also compare our results with observational constrains for SMBHs and wind-fed SgXBs and find overall good agreement (Section \ref{sec:obs}). Our main findings include:
\begin{itemize}
    \item Turbulence in SCAFs is driven locally by convection, as opposed to injected by the accretor or the large-scale flow (Section \ref{sec:turbulence}).
    
    \item The radial scalings of SCAFs are set by radial momentum balance, which yields (Section \ref{sec:scaling})
    \begin{align}
        &\alpha_\rho \equiv \frac{{\rm d}\log\rho}{{\rm d}\log r} = -1\pm{\rm few}~0.1,\\
        &\alpha_{\dot M} \equiv \frac{{\rm d}\log\dot M}{{\rm d}\log r_{\rm acc}} = 0.5\pm{\rm few}~0.1.
    \end{align}
    The uncertainties are because our argument relies upon the up-down symmetry of the turbulence, which is only approximate for large-scale turbulence. Empirically, for an adiabatic hydrodynamic flow with $\gamma=5/3$, we get
    \begin{equation}
        \alpha_\rho\approx -0.8, ~~~\alpha_{\dot M}\approx 0.7,
    \end{equation}
    which agrees well with the observed accretion rates for SMBHs and wind-fed SgXBs (Fig. \ref{fig:obs_Mdot}).

    \item Local turbulence in the accretion flow leads to a pink-noise variability, which appears consistent with the long-term variability of Sgr A* (Fig. \ref{fig:obs_SgrA}).

    \item Theoretically, we expect the above properties of SCAFs to be relatively robust and insensitive to the inclusion of additional physics such as cooling and magnetic field; this explains the striking similarity between simulations with different physical setups (Section \ref{sec:discussion:robusity}). We further conjecture that hot MADs exhibiting strong turbulence driven by a convection-like instability \citep{Begelman2022} may be interpreted as a special type of SCAF (Section \ref{sec:MAD}).

    \item SCAFs and CDAFs correspond to two distinct regimes of convective accretion flows. Qualitatively, SCAF/CDAF is preferred if angular-momentum transport by convection is outward/inward (Section \ref{sec:dis:disk}). Empirically, 3D simulations with $\gamma\approx5/3$ generally show SCAFs.
\end{itemize}

The simple scalings of SCAFs, together with their robustness, produce a wide range of potential applications.
When modeling systems with a small accretor globally, the $\dot M$ -- $r_{\rm acc}$ scaling provides a simple model of accretion (and feedback) when directly resolving $r_{\rm acc}$ is unrealistic.
On the other hand, when studying the dynamics near the accretor (e.g., modeling horizon-scale observations of SMBHs), the SCAF solution serves as an alternative initial condition and/or outer boundary condition that might be more realistic in some scenarios.

~

We thank James Stone for valuable comments on an early draft and helpful suggestions regarding the comparison with CDAF.
We also appreciate helpful discussions with
Christopher White,
Minghao Guo,
Mordecai-Mark Mac Low,
Phillip Armitage,
Yan-Fei Jiang,
Yixian Chen,
and many others, as well as insightful comments from the anonymous referees.
Simulations in this work are performed with computational resources at the Flatiron Institute.
\software{
\texttt{athena++} \citep{Stone2020}
}

\appendix
\section{A model for mean radial acceleration at finite mixing length}
\label{a:finite_xi}

In this appendix we construct a simple model of convective turbulence with finite mixing length $\xi\sim r$ and use it to argue that zero net mass flux requires small mean acceleration and $\alpha_\rho\approx -1$.

In general, we can separate the radial acceleration in the flow into the following three components: First, a convective acceleration that is proportional to the adiabatic displacement the current parcel of gas has traveled through. Suppose this displacement is $\Delta\log r = x$, the convective acceleration is
\begin{equation}
    a_{\rm conv} \equiv a_{P}(e^{fx}-1) \approx fa_P x \left(1+\frac 12 fx\right).
\end{equation}
Here $a_{P}\equiv-\bar\rho^{-1}{\rm d}\bar P/{\rm d}r$ is the pressure support, and $f \equiv -N^2r/a_P$ corresponds to the power-law slope of how $\bar\rho/\rho$ scales with displacement. $f$ is usually small; for example, for $\gamma=5/3$ and $\alpha=-1$, we have $f=1/5$. We keep terms up to $\mathcal O(x)$ relative to the lowest order term, because it will eventually affect the $\mathcal O(1)$ term of $\alpha_\rho+1$.
For the remaining acceleration, we can separate it into a constant background acceleration and a turbulent acceleration,
\begin{equation}
    a_{\rm bg} \equiv \bar\rho^{-1}\overline{\rho(a_r - a_{\rm conv})},~~~a_{\rm turb} \equiv a_r-a_{\rm conv}-a_{\rm bg}.
\end{equation}
For simplicity, we assume that $a_{\rm turb}$ on average does not affect the radial velocity, so we can estimate the mean radial velocity at given displacement using $a_{\rm conv}$ and $a_{\rm bg}$ alone.
One caveat is that we make this assumption mainly because modeling the effect of $a_{\rm turb}$ more realistically is too challenging, and we do do not have a rigorous argument on why (or whether) this is generally a reasonable approximation.

Now consider the radial velocity and acceleration at a given radius $r$.
The flow consists of parcels that traveled from different initial locations; for a parcel that originates from $r'$, with displacement $x=\log(r/r')$, the velocity is given by (ignoring the contribution from $a_{\rm turb}$)
\begin{equation}
\begin{split}
    \frac 12 v_r^2=&\int_{r'}^r [a_{\rm conv}(r'')+a_{\rm bg}(r'')]{\rm d}r''\\
    =& r a_{P}\int_0^x (e^{fx'}-1)e^{x-x'}{\rm d}x' + r a_{\rm bg}\int_0^x e^{x-x'}{\rm d}x'\\
    \approx & r fa_P\left[\frac 12 x^2 + \left(\frac 16 + \frac 16 f\right) x^3\right] + ra_{\rm bg}x;
\end{split}
\end{equation}
\begin{equation}
    v_r \approx \sqrt{rfa_P} \left[x + \left(\frac 16 + \frac 16 f\right) x^2 + \frac{a_{\rm bg}}{fa_P}\right].\label{eq:v}
\end{equation}
Here all variables are evaluated at $x$ unless otherwise noted. We keep terms up to $\mathcal O(x)$ relative to the lowest order term, in order to give a basic estimate of the effect of finite $x$ and $a_{\rm bg}$.
Eq. \eqref{eq:v} allows us to write the acceleration in the following form,
\begin{equation}
    a_{\rm conv}+a_{\rm bg} \approx v_r\sqrt{r^{-1}fa_P} + \left(\frac 13 f - \frac 16\right) f a_P x^2.
    \label{eq:a_in_v}
\end{equation}
Now consider the mean acceleration, which is just a density-weighted average of $a_{\rm conv}+a_{\rm bg}$ across all parcels at given radius. (By definition, mean $a_{\rm turb}$ is zero.) Since zero net mass flux implies $\overline{\rho v_r}=0$, the first term in Eq. \eqref{eq:a_in_v} averages to zero, and
\begin{equation}
    \bar\rho a_{r,0} \approx \left(\frac 13 f - \frac 16\right) f a_P \overline{\rho x^2} \approx \left(\frac 13 f - \frac 16\right) \frac{\overline{\rho v_r^2}}{r}.
\end{equation}
Then, using Eq. \eqref{eq:alpha_in_a_r}, $\alpha_\rho$ is given by
\begin{equation}
    \alpha_\rho + 1 = \left(\frac 13 f - \frac 16\right) + \mathcal O(x). \label{eq:alpha_est}
\end{equation}

For $f=1/5$ (corresponding to $\gamma=5/3$ and $\alpha_\rho=-1$), the first term in the RHS of Eq. \eqref{eq:alpha_est} is $-0.1$. The amplitude of this term remains similarly small when we vary $\gamma$, as $f$ remains small and positive. The second $\mathcal O(x)$ term may become comparable to the first term due to the finite mixing length, but we do not expect it to be significantly larger than the first term. This is mainly because the prefactor of this term originates from $\mathcal O(x^4)$ terms in $v_r^2$ and is therefore small (expanding exponential functions to order $n$ gives a prefactor of $\sim 1/n!$).
We also comment that the typical value of $x$ is not too large; since the mixing length should be comparable to the density scale height \citep{Arnett2018}, the typical value of $x$, which is half of the mixing length, is $x\sim 1/(2|\alpha_\rho|)\sim 1/2$.
In summary, while we cannot obtain an accurate estimate of $\alpha_\rho$ because the high-order terms in Eq. \eqref{eq:alpha_est} can have finite amplitude, the small amplitude of the zeroth order term (and similarly small prefactors for higher order terms) suggests that $\alpha_\rho+1$ is likely small and of order few 0.1.

We comment that this analysis comes from a somewhat different perspective compared to the up-down symmetry argument in the main text. In this analysis, we make no assumption at all about how symmetric the turbulent motion has to be. Instead, we only argue that $a_r$ is close to being linearly proprotional to $v_r$ when we remove the turbulent part of the dynamics, and therefore zero mass flux (zero $v_{r,0}$) corresponds to small $a_{r,0}$.

\section{Observational estimates of normalized accretion rate and accretor size}\label{a:obs}

In this appendix we discuss how the normalized accretion rates and accretor sizes in Fig. \ref{fig:obs_Mdot} are estimated.

For M87*, we take a black hole mass of $M_{\rm BH} = 6.5\pm 0.7\times 10^9 {\rm M}_\odot$ \citep{EHT2019}, and an accretion rate of $\dot M = 3-20 \times 10^{-4} {\rm M}_\odot/{\rm yr}$ \citep{EHT2021}. The Bondi radius and Bondi accretion rate are computed based on the electron density and temperature measurements in \citet[][see their Section 3.4]{Russell2015}.

For Sgr A*, we take a black hole mass of $M_{\rm BH} = 4.0_{-0.6}^{+1.1}\times 10^6{\rm M}_\odot$ \citep{EHT2022a} and an accretion rate of $\dot M = 5.2-9.5 \times 10^{-9} {\rm M}_\odot/{\rm yr}$ \citep{EHT2022b}. The Bondi radius and Bondi accretion rate are taken from the simulation in \citet[][last paragraph of Section 4.3.1]{Ressler2018}, which simulates the accretion of the wind launched by $\sim 30$ Wolf-Rayet stars within the central parsec of the galactic center onto Sgr A*. The simulation takes $M_{\rm BH} = 4\times 10^6{\rm M}_\odot$, consistent with more recent observations, and the properties of the stars, including their wind launching rates and wind velocities, are directly taken from observations. This simulation includes \revise{essentially} no free parameter and has not been tuned to match any observation of the accretion flow.

Data for the wind-fed SgXBs are taken from Fig. 28 of \citet{XuStone2019}, and below we briefly discuss how the normalized accretion rate and accretor size are defined. In a wind-fed SgXB system, we have a NS accreting from the supersonic wind of a supergiant companion. Laminar accretion from a supersonic inflow with density $\rho_\infty$ and velocity $v_\infty$ can be described by Bondi-Hoyle-Lyttleton accretion, where the flow forms a bow shock in front of the accretor and the flow behind the shock is similar to a Bondi flow (see a review in \citealt{Edgar2004}). This gives a characteristic accretion radius ($\sim$ shock radius) and accretion rate
\begin{equation}
    R_{\rm a} \equiv \frac{2GM}{v_\infty^2}, ~~~ \dot M_{\rm HL} \equiv \pi R_a^2\rho_\infty v_\infty.
\end{equation}
They are analogous to the Bondi radius and Bondi accretion rate.
We use $R_{\rm a}$ and $\dot M_{\rm HL}$ to normalize the accretor radii and accretion rate for SgXBs in Fig. \ref{fig:obs_Mdot}. For the accretor radius $r_{\rm acc}$, we take the magetosphere radius as this is roughly where material begins to be captured by the NS. Details on how relevant system parameters are estimated are presented in \citet[][Section 2 and 7.2]{XuStone2019}; the uncertainties in these parameters, especially the assumed magetosphere radius, are hard to estimate and can be quite large.

We note in passing that \citet{XuStone2019} predict stable (laminar) accretion flow for one system in this sample, 4U 1907+097, but that prediction may be incorrect.
The stability threshold in \citet{XuStone2019} is obtained from simulations with a vacuum sink, which tend to be more stable for large $r_{\rm acc}$ than a smooth sink (cf. Fig. \ref{fig:Mdot_and_rho_scaling}). Meanwhile, a smooth sink might better resemble the accretion onto magentospheres, and if we infer stability based on the results of smooth sink simulations, all the observed systems considered in \citet{XuStone2019} probably have turbulent flows given their small $r_{\rm acc}/R_{\rm a}$.

\bibliography{Xu23}{}
\bibliographystyle{aasjournal}

\end{document}